\documentclass[aps,prl, twocolumn, amsmath,amssymp,showpacs]{revtex4-1}

\usepackage[T1]{fontenc}
\usepackage[latin1]{inputenc}
\usepackage[dvips]{graphicx,color}
\usepackage{rotating}
\usepackage{bm}        
\usepackage{amssymb}   
\usepackage{amsmath} 
\usepackage{bm}
\def\jcp#1#2#3{J.~Chem.~Phys.~{\bf #1},\ #2\ (#3)}
\def\cpl#1#2#3{Chem.~Phys.~Lett.~{\bf #1},\ #2\ (#3)}

\def\pra#1#2#3{Phys.~Rev.~A~{\bf #1},\ #2\ (#3)}
\def\prl#1#2#3{Phys.~Rev.~Lett.~{\bf #1},\ #2\ (#3)}

\def\ii{\text{i}}

\def\k1{k_1}
\def\k2{k_2}
\def\q1{q_1}
\def\q2{q_2}

\def\({\left (}
\def\){\right )}
\def\[{\left [}
\def\]{\right ]}

\newcommand{\beq}{\begin{equation}}
\newcommand{\eeq}{\end{equation}}

\begin{document}
\date{\today}
\flushbottom \draft
\title{Long-lived quasi-stationary coherences in V-type system driven by incoherent light} 

\author{Timur V. Tscherbul and Paul Brumer}
\affiliation{Chemical Physics Theory Group, Department of Chemistry, and Center for Quantum Information and Quantum Control, University of Toronto, Toronto, Ontario, M5S 3H6, Canada} \email[]{ttscherb@chem.utoronto.ca}

\begin{abstract}
We present a theoretical study of noise-induced quantum coherences in a model three-level V-type system interacting with incoherent radiation, an important prototype for a wide range of physical systems ranging from trapped ions  to biomolecules and quantum dots. 
By solving the quantum optical equations of motion, we obtain {\it analytic} expressions for the noise-induced coherences and show that they exhibit an oscillating  behavior in the limit of large excited level spacing $\Delta$ ($\Delta /\gamma \gg 1$, where $\gamma$ is the radiative decay width).  Most remarkably, we find that in the 
opposite limit of small level spacing $\Delta/\gamma \ll 1$, appropriate for large molecules, (a) the coherences can survive for an extremely long time $\tau=(2/\gamma) (\Delta/\gamma)^{-2}$ before eventually decaying to zero, and (b) coherences at short times can be substantial. We further show that  the long-lived coherences can 
survive environmental relaxation and decoherence, suggesting implications to the design of quantum heat engines and to incoherent light excitation of biological systems.
\end{abstract}

\maketitle
\clearpage
\newpage

The dynamics of atomic and molecular systems interacting with noisy electromagnetic fields (such as blackbody radiation) is a recurring theme of interest in physics, chemistry, and biology.  
In particular, the possibility of generating long-lived  quantum mechanical coherence via the interaction of multilevel atomic and molecular systems with incoherent light has recently attracted much interest \cite{Scully11,Scully13,Scully06,Cam13}.  
Apart from their fundamental importance,  such noise-induced coherences have a wide range of proposed applications, ranging from enhancing the efficiency of photovoltaic devices \cite{Scully11} to lasers without inversion \cite{Scully06} and to the design of artificial light-harvesting antenna complexes \cite{Scully13}.  The noise-induced coherences are created via quantum (Fano) interference of the transition amplitudes leading to the same final state---processes similar to those giving rise to well-known coherent optical phenomena such as coherent population trapping \cite{CPT}, electromagnetically induced transparency \cite{EIT}, vacuum-induced coherence \cite{VIC1,VIC2,VIC2a}, and coherent control \cite{SBbook}.




Previous theoretical studies have explored the properties of noise-induced coherences in atomic $\Lambda$ and V-type systems \cite{F92,Altenmuller,Scully06,Ou,Plenio} and  
 suggested their role in enhancing the efficiency of photosynthetic energy transfer in light-harvesting antenna complexes \cite{Scully13}.
However, this work  focused on the 
analysis of steady-state properties \cite{Scully06,Ou,Scully11,Scully13} and 
paid little attention to the time evolution of the coherences. 
In an earlier contribution, Hegerfeldt and Plenio \cite{Plenio} considered the dynamics of a three-level atomic ion pumped by incoherent radiation and focused on the nature of the emitted light. They demonstrated that in the limit of large excited-state splitting
the intensity correlation function exhibits quantum beat oscillations due to noise-induced coherences, and in the opposite limit the coherences last longer, leading to extended dark periods in the emitted light. However,  no closed analytic expressions were derived  for the noise-induced coherences, so their time scale and the influence of various system parameters (such as the excited level splitting, and the radiative decay rates) remained unexplored, making it difficult to appreciate and to experimentally observe these coherences in real  atomic and molecular systems.

  In this Letter, we focus on the dynamical properties of the V-type system, a minimal ``building block'' for quantum heat engines based on Fano interference \cite{Scully13}, crucial to understanding the phenomena of incoherent light excitation in visual phototransduction and solar light harvesting. 
   Our analysis is based on a quantum optical master equation of non-Lindblad type \cite{F92,Scully06} that treats all density matrix elements (state populations and coherences) on an equal footing.  {This treatment goes beyond the traditional secular approximation, which underlies the standard rate equation formalism for incoherent light excitation of molecular systems \cite{CohenTannoudji} but breaks down for nearly degenerate states of the V-type system  \cite{CohenTannoudji,Plenio},
     leading to a more complex master equation in which populations and coherences are inextricably coupled.} 
    We find analytic solutions to this equation in two 
  opposite limits and show that the populations remain positive at all times, thereby demonstrating the existence of positive-definite solutions of a non-Lindblad-type master equation. Our analytic results show that when the energy splitting $\Delta$ between the two upper levels of the V-type system  is large compared to the radiative decay widths $\gamma$, the off-diagonal elements of the density matrix exhibit coherent oscillations that decay on the timescale $\tau_s = 1/\gamma$ 
  (hereafter we set $\hbar =1$).
We  further  show  that  in the limit of small energy splitting
between   excited-state   levels   ($\Delta\ll   \gamma$),  (a)
short-time  coherences  are  significant, and (b) the coherence
between  the  upper  energy  levels  survives  on  a  timescale
$\tau_\text{long}  = \frac{2}{\gamma}(\Delta/\gamma)^{-2}$ that
by  far  exceeds  any  other dynamical timescale in the problem
(including  that  of spontaneous decay), even in the presence of decoherence and relaxation. 
This remarkable result
suggests that coherent dynamics in V-type molecular systems can
survive  for  very long  times,   with immediate
implications  for the variety of applications cited above.

To elucidate the time dynamics of the populations and coherences,  we solve the Born-Markov quantum optical master equation \cite{BPbook} in the energy basis
\begin{align}\notag
\dot{\rho}_{ii} &= -(r_i+\gamma_i +\Gamma_i)\rho_{ii} + r_i\rho_{cc} - p (\sqrt{r_a r_b} + \sqrt{\gamma_a\gamma_b}) \rho_{ab}^R  \\ \label{EOMgeneral} 
\dot{\rho}_{ab} &= -\frac{1}{2}(r_a+r_b +\gamma_a + \gamma_b +2\gamma_d)\rho_{ab} - \ii\rho_{ab}\Delta \\ &+ \frac{p}{2}\sqrt{r_a r_b}(2\rho_{cc} - \rho_{aa}-\rho_{bb}) \notag
- \frac{p}{2}\sqrt{\gamma_a\gamma_b}(\rho_{aa}+\rho_{bb}),
\end{align}
where $\rho_{ab}=\rho_{ab}^R + \ii \rho_{ab}^I$, $\gamma_i$ is the radiative width of level $|i\rangle$, $i=a,b$ (see 
Fig. 1a), $r_i$ is the incoherent pumping rate, 
and $p=\bm{\mu}_{ac}\cdot \bm{\mu}_{bc}/(\mu_{ac}\mu_{bc})$ quantifies the angle between the $c\to a$ and $c\to b$ transition dipole moments  (below we  focus on the case $p=1$,
as justified in the Supplementary Material\cite{SI}). In Eqs. (\ref{EOMgeneral}), $\Gamma_i$ are the (phenomenological) relaxation rates, and $\gamma_d$ is the decoherence rate. For a molecule in the absence of an external environment, $\Gamma_i=\gamma_d=0$.
Following previous theoretical work \cite{Scully06}, we adopt $r_a=r_b=r$, $\gamma_a = \gamma_b = \gamma$. This assumption greatly simplifies the analytical solution of Eqs. (\ref{EOMgeneral}) while 
retaining the essential physics of the problem \cite{note1}. 
We note that in deriving Eqs. (\ref{EOMgeneral}), we do not make the secular approximation,  {which neglects the elements of the relaxation tensor $R_{ijkl}$ \cite{CohenTannoudji} with $\omega_{ij} \ne \omega_{kl}$, where $\omega_{ij}$ is the 
energy gap $\epsilon_i-\epsilon_j$ between eigenstates $|i\rangle$ and $|j\rangle$ . This procedure is commonly justified by noting that the oscillatory factors $e^{\ii (\omega_{ij}-\omega_{kl})t}$ multiplying $R_{ijkl}$ average to zero if the timescale of interest is much longer than the system evolution time $1/\omega_{ij}$. The secular approximation breaks down for a V-type system where $\omega_{ab}=\Delta > 0$ can be arbitrarily small.} However, the inclusion of non-secular terms means that 
Eqs.~(\ref{EOMgeneral}) are no longer of Lindblad type
\cite{BPbook}, and can therefore have spurious solutions where state populations $\rho_{ii}(t)$ can take unphysical negative values \cite{Silbey92}.  The use of non-Lindblad-type master equations to describe open quantum  systems has  been the subject of  ongoing debate \cite{Cam13}. Fortunately, as  shown here, the existence of positive solutions  is guaranteed for a V-type system subject to initial conditions appropriate to weak-field incoherent excitation.

\begin{figure}[t]
	\centering
	\includegraphics[width=0.35\textwidth, trim = 0 0 0 0]{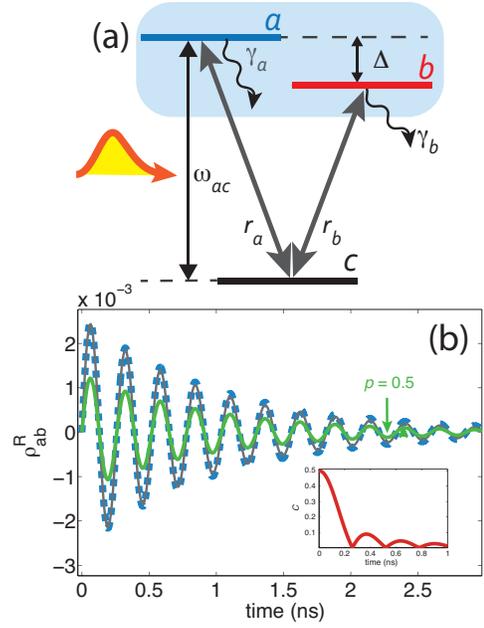}
	\renewcommand{\figurename}{Fig.}
	\caption{(a) The energy level structure of the V-type system. The shaded area represents the effects of environment-induced relaxation and decoherence on the excited states $a$ and $b$ with $\Delta=\epsilon_a - \epsilon_b$.  The real part of the two-photon coherence $\rho_{ab}^R(t)$ (b) and the ratio $\mathcal{C}=|\rho_{ab}|/(\rho_{aa}+\rho_{bb})$ (inset).  Full lines -- exact result (numerical calculations), dashed lines -- analytical result obtained in this work in the $\Delta/\gamma \gg 1$ limit ($\gamma/2\pi$ = 1 GHz, $\Delta/\gamma = 24$).  The incoherent pumping rate $r/\gamma=\bar{n}= 0.063$, where $\bar{n}=1/[e^{\hbar\omega_{ac}/k_BT}-1]$ is the thermal occupation number ($k_BT=0.5$ eV for sunlight), and $\omega_{ac}=1.41$~eV. 
	Results for $p= 1/2$ are also shown (green/light grey line).}\label{fig:1}
\end{figure}

The solution of Eqs. (\ref{EOMgeneral}) can be represented as $\rho_{aa}(t) = -a/a_0 + \sum_i {\tilde{c}}_i e^{\lambda_i t}$,
where  $\lambda_i$ are the roots of the cubic polynomial $\lambda^3 + a_2 \lambda^2 + a_1 \lambda + a_0 = 0$, and $\tilde{c}_i$ are time-independent coefficients determined from initial conditions  {appropriate to incoherent excitation from the ground state, [$\rho_{ij}(0)=0$, $\rho_{cc}(0)=1$]} (all results below are derived in the Supplementary Material \cite{SI}).
 The {\it exact} time dynamics of the coherence between the excited-state levels (hereafter referred to as simply ``coherence'')  is 
$\rho_{ab}^R(t) = -\frac{1}{p(r+\gamma)} \sum_i \tilde{c}_i (3r+\gamma+\lambda_i) e^{\lambda_i t}$.
The  behavior  of  the  coherences  in  a V-type system is thus
determined  by  the  values  of  the roots $\lambda_i$. Two 
important physical regions can be identified: 
$\Delta/\gamma \gg 1$ (one real and two complex conjugate $\lambda_i$),
and $\Delta/\gamma \ll 1$ (real $\lambda_i$). 

Consider first the limit of large excited-state splitting, $\Delta/\gamma \gg 1$, applicable to weak-field ($r/\gamma \ll1$) incoherent excitation \cite{Hoki,Zaheen} of small to medium-sized molecules with typical  excited-state lifetimes of $1-10$ ns and the density of states of less than 1 state in $\gamma \sim 0.005 - 0.05$ cm$^{-1}$. Under these conditions, the roots of the cubic equation are $\lambda_1 = -\gamma$, $\lambda_{2,3}=-\gamma\pm \ii\Delta$, and the upper-state population and the real part of the coherence take the form 
\begin{align}\label{coh_largeDelta}
\rho_{aa}(t) &= \left( {r}/{\gamma}\right) \left[ 1 - e^{-\gamma t}  \right], \\
\rho_{ab}^R(t) &= \left( {r}/{\Delta}\right) e^{-\gamma t} \sin (\Delta t) \qquad (\Delta/\gamma \gg 1), \label{pop_largeDelta}
\end{align}
with the imaginary part of the coherence $\rho_{ab}^I = ( \frac{r}{\Delta}) e^{-\gamma t} [\cos(\Delta t) -1]$. 
Figure 1(b) confirms that the analytic result (\ref{coh_largeDelta}) is in excellent agreement with the exact time evolution of $\rho_{ab}^R(t)$ obtained by numerical 
integration of Eq. (\ref{EOMgeneral}), here for $\Delta/\gamma = 24$. The coherence exhibits damped oscillations with frequency $\Delta$ and 
decays to zero on the timescale $\tau_s = 1/\gamma$, {\it i.e.} the decoherence time scale is given here
by the radiative lifetime of the excited-state levels $\tau_s = 1/\gamma$. On short timescales ($t\ll \tau_s$), spontaneous decay can be neglected, and the decoherence timescale, quantified by the ratio $\mathcal{C}=|\rho_{ab}|/(\rho_{aa}+\rho_{bb})$ \cite{PRA,Zaheen},  decreases as $1/|\Delta t|$ \cite{PRA}. This dependence reflects the linear growth of excited-state population (Eq. (\ref{coh_largeDelta}) yields $\rho_{aa} = rt$ in the limit $t\ll 1/ \gamma$), while the coherences are bounded by $r/\Delta$.  The 
inset of Fig.~1(b) illustrates the slow decrease of $\mathcal{C}(t)$, which is  consistent with our previous  results obtained independently via a different theoretical approach  \cite{JiangBrumer,Zaheen,PRA}. However, Eqs.~(\ref{coh_largeDelta}) and (\ref{pop_largeDelta})  are required for longer timescales.  {As shown in Fig. 1(b), reducing $p$  leads to a decrease of the coherence without affecting the overall time dynamics. }

\begin{figure}[t]
	\centering
	\includegraphics[width=0.57\textwidth, trim = 130 40 0 0]{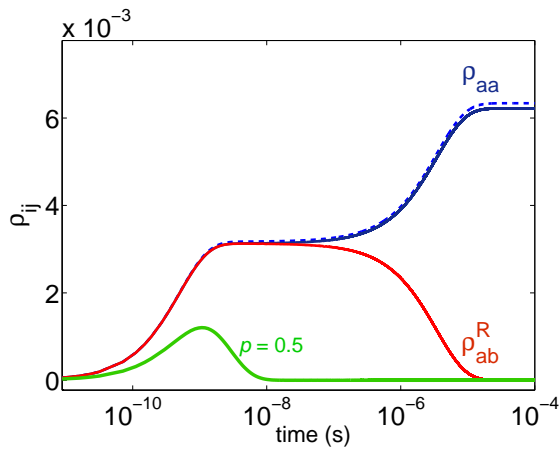}
	\renewcommand{\figurename}{Fig.}
	\caption{Excited-state population and coherence dynamics  in the $\Delta/\gamma \ll 1$ limit  ($\Delta/\gamma = 0.024$, $\gamma/2\pi=1$~GHz, $r/\gamma=0.1 \bar{n}$). Exact results (full line), analytic results (dashed line), results for $p=1/2$ (green/light grey line)}\label{fig:2}
\end{figure}

In the opposite limit $\Delta/\gamma \ll 1$, which applies to incoherent excitation of a V-type system with very closely spaced upper levels ({\it e.g.} a 
polyatomic molecule like Antracene has $\Delta/\gamma \sim 10^{-7}$ \cite{Antracene}),  the roots of the cubic equation are given by $\lambda_1 = -2\gamma$, $\lambda_2=-\gamma$, and $\lambda_3 = -\frac{1}{2}\gamma (\Delta/\gamma)^2$. For the upper-state population and the real part of the coherence we find in the weak pump limit $ r/\gamma \ll 1$ ($r/\gamma\sim10^{-9}$ for sunlight illumination \cite{Hoki})
\begin{align}\label{pop_smallDelta}
\rho_{aa}(t) &=  \left( {r}/{2\gamma}\right) \left[ 2 - e^{-2\gamma t} - e^{-\frac{1}{2}\gamma (\Delta/\gamma)^2t}     \right],  \\	\label{coh_smallDelta}
\rho_{ab}^R(t) &= \left( {r}/{2\gamma} \right) \left[ e^{-(\gamma/2) (\Delta/\gamma)^2 t} - e^{-2\gamma t}\right]  .							
\end{align}
whereas the imaginary part of the coherence $\rho_{ab}^I = ( \frac{r\Delta}{2\gamma^2}) [ 2e^{-\gamma t} - e^{-2\gamma t} - e^{-\frac{1}{2}\gamma (\Delta/\gamma)^2t} ]$, 
is smaller by the factor $(\Delta/\gamma) \ll 1$.  
We observe that in the $\Delta/\gamma\to 0$ limit, the timescale for the decay of the first exponent in Eq. (\ref{coh_smallDelta}), $\tau_\text{long} = (2/\gamma) (\Delta/\gamma)^{-2}$ approaches infinity. Thus, Eq.~(\ref{coh_smallDelta}) establishes the existence of two widely disparate timescales in coherence dynamics: At shorter times ($t\sim 1/\gamma$), the first exponent on the right-hand side is close to unity, so the coherence grows from zero to a quasi-steady-state value  $r/2\gamma$. At $t\gg 1/\gamma$ but still short compared to $\tau_\text{long}$, the coherence remains constant, before eventually decaying to zero at $t > \tau_\text{long}$.

To  illustrate this behavior, we provide a log plot in Fig.~2 of the time variation of the population $\rho_{aa}(t)$ and of the coherence $\rho_{ab}^R$ for $\Delta = 10^{-7}$ eV, in the $\Delta/\gamma \ll 1$ limit. The extremely long survival time of the quasi-stationary coherence is apparent: for the given parameters, it survives for as long as 10 $\mu$s, which is $\sim$10$^4$ times longer than the excited states' radiative lifetime. This is a consequence of the continuous pumping and the resultant establishment of a quasi-steady state. 
\textit{In the limit of $\Delta \to 0$, the timescale for the existence of the quasi-stationary coherence approaches infinity}, underscoring the crucial role of the excited-state level splitting $\Delta$ in determining the noise-induced coherence dynamics. Setting $\Delta = 0$ in Eq. (\ref{coh_smallDelta}) produces a non-zero  steady-state value of the coherence $\rho_{ab}^R(\infty)=r/2\gamma$, 
in agreement with prior $\Delta=0$ results \cite{Scully06}. 
However, for any $\Delta > 0$ the coherences  have a finite lifetime $\tau_\text{long} = (2/\gamma) (\Delta/\gamma)^{-2}$. 
Maximizing the lifetime of the coherences  is proposed to be beneficial, {\it e.g.}, to the design of
quantum heat engines based on Fano interference \cite{Scully11,Scully13},
and our analysis suggests the benefit of using as small a $\Delta$ as possible.
As shown in Fig. 2, changing $p$ from 1 to $1/2$ causes the coherence to decrease and shortens its lifetime.

Both the analytical prediction (\ref{coh_smallDelta}) and the numerical results for $p=1$  establish that the time  dependence of excited-state populations closely follows that of the coherences up until $t=\tau_s$, thereby implying significant short-time coherence effects, with $\mathcal{C}(t) = 1/2$. 
We emphasize that this behavior is drastically different from the $1/|t\Delta|$ decay of the coherence-to-population ratio in the limit $\Delta/\gamma \gg 1$ shown in the inset of Fig. 1(b). {\it This indicates that very closely spaced energy levels 
maintain coherence on a much longer timescale than levels separated by an energy gap that is large compared to their natural linewidth. }
Finally, we note that in both small-$\Delta$ and large-$\Delta$ regimes, the populations $\rho_{ii}(t)$ remain positive at all times in the $r/\gamma \ll 1$ limit.

  \begin{figure}[t]
	\centering
	\includegraphics[width=0.35\textwidth, trim = 0 0 0 0]{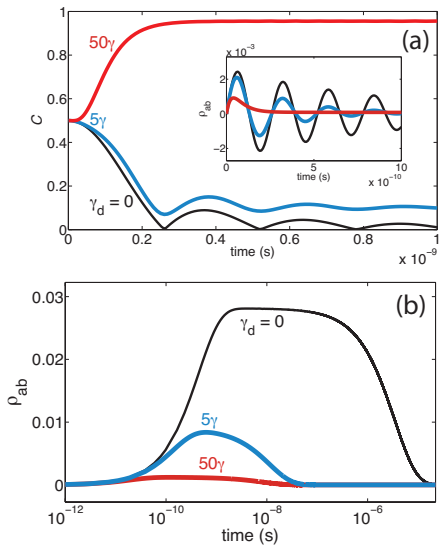}
	\renewcommand{\figurename}{Fig.}
	\caption{Effects of environmental relaxation and decoherence on V-type system dynamics:  (a) The $\mathcal{C}$-ratio and the real part of the coherence (inset) 
	as a function of time for $\Delta/\gamma=24$ and $\gamma_d/\gamma=$ 0, 5, 50.  The decoherence rates $\gamma_d$ are shown next to each curve. (b) The real part of the coherence for $\Delta/\gamma=0.024$ and $\gamma_d/\gamma=0$,  5, 50. The imaginary part of the coherence is negligible compared to the real part.}\label{fig:3}
\end{figure}

{\it Effects of Relaxation and Decoherence}. Thus far we have considered an isolated V-type system. 
However, the excited states of a system within a condensed-phase environment ({\it e.g.} a molecule in a liquid) are subject to relaxation and decoherence. 
To model these effects, we consider nonzero $\Gamma$ and $\gamma_d$ in Eqs.~(\ref{EOMgeneral}), and modify the equations to include relaxation to an auxiliary level $|d\rangle$.
 Figure 3(a) illustrates the effect of environmental relaxation and decoherence
 in the large $\Delta/\gamma$ regime using the $\mathcal{C}$-ratio, which quantifies the amount of coherences relative to populations. We observe that for  moderate decoherence ($\gamma_d=5\gamma$) the ratio displays periodic oscillations superimposed on a slowly decaying background. This time dependence is similar to that observed in the absence of relaxation and decoherence; however, the asymptotic value of $\mathcal{C}$ does not decay to zero due to a finite steady-state value of $|\rho_{ab}|$ (not shown).

 The origin of this counterintuitive  behavior can be understood by noting that relaxation prevents excited-state population from accumulating by causing decay to the  auxiliary level $|d\rangle$. While the coherences are also suppressed by environmental decoherence, this effect is not as significant as relaxation-induced suppression of excited-state populations. As a result, the populations can no longer outgrow the coherences and the value of $\mathcal{C}$ saturates at a steady state.
  As shown in the inset of Fig.~3(a),  the value of the decoherence rate $\gamma_d$ has a major effect 
  on the dynamics in the $\Delta/\gamma \gg 1$ regime, suppressing the
  coherent oscillations.  Already for $\gamma_d = 5\gamma$, the oscillations are damped on the timescale $\tau_d = 1/\gamma_d$, {\it i.e.} much sooner than in the absence of environment ($\tau_s=1/\gamma$), indicating that in the regime $\gamma_d\gg \gamma$, the decoherence timescale is set by the environment-induced decoherence, rather than spontaneous decay.  
 
Figure 3(b) shows that  relaxation  and decoherence lead to a
suppression  of  the quasi-stationary  coherences in the $\Delta/\gamma \ll 1$ regime,  and  cause a decrease in the
intermediate  ``plateau'' values reached at $t\gg 1/\gamma$. In
addition,  the   time  of reaching the  plateau 
decreases  with  increasing  $\gamma_d$, indicating that in the $\gamma_d \gg \gamma$ regime,
the dynamics are governed
by  a  timescale  $\tau_d  = 1/\gamma_d$. Nevertheless,
the coherences are still substantial in magnitude, and
the  $\mathcal{C}$-ratio  (not shown) remains  close  to unity over a large
time interval even in the presence of relaxation and decoherence\cite{SI}.
  


To begin to explore the applicability  of these results to real molecules, we solved the quantum optical master equation for a multilevel analog of the V-type system \cite{SI}. The multilevel  excited-state coherences were found to qualitatively follow the
 behavior observed in Figs. 1 to 3, and the coherences to persist relative to the populations.  
The V-type system thus provides insight into multilevel dynamics, a consequence of the weak intensity of the incoherent light and the absence of direct light-induced coupling between the excited states \cite{SI}.


 {Theoretical work on atomic systems \cite{F92,VIC2} and quantum dots  \cite{Scully06} has shown that the key condition for the generation of  noise-induced coherences---the existence of non-orthogonal  transition dipole moments ($p\ne 0$)---requires two closely spaced excited levels with the same values of total angular momentum $J'$ and its space-fixed projection $M'$. 
 The optimal conditions of $p=1$ can be realized, {\it e.g.} for the closely spaced Rydberg states \cite{PRA}, or  excited vibrational states of polyatomic molecules with the same  $J'$ and $M'$ (see  Supplementary Material \cite{SI}).}


In summary, we have presented a theoretical analysis of noise-induced coherences in a V-type molecular system interacting with incoherent radiation, a basic model for a wide array of excitation processes in atomic, molecular, and solid-state systems.  
This work demonstrates that, contrary to expectation,  incoherent light 
excitation  can generate long-lived quasi-stationary coherences that can survive  
environment-induced decoherence, and display considerable coherence at short time.
 As a consequence,
these results strongly motivate examination of their role in
  biological  processes, such as cis-trans photoisomerization of
  retinal\cite{TBretinal}  (the  first step in vision) or photosynthetic energy
  transfer.  
  

We thank Profs. Leonardo Pach\'on and Robert Field for discussions. This work was supported by NSERC of Canada and by the U.S. Air Force Office of Scientific Research under contract number FA9550-13-1-0005.



\begin{thebibliography}{99}
\bibitem{note1}
We  verified by explicit numerical solution of Eqs. (\ref{EOMgeneral}) that the density matrix dynamics obtained for $r_a \ne r_b$ is very similar to that observed for $r_a=r_b$ provided that the pumping rates are within a factor of 2 of one another.
\end{thebibliography}
\end{document}